\documentclass[11pt]{article}
\usepackage{amssymb,palatino,lscape,a4wide,subfigure,graphicx}

\begin{document}
\topmargin -0.9in
\textheight 9.5in

\newcommand{\ea}{et al.}
\newcommand{\be}{\begin{equation}}
\newcommand{\ee}{\end{equation}}
\baselineskip 21pt
\parindent 10pt
\parskip 9pt

\noindent {\large\bf London house prices are power-law distributed}\footnote{second version with minor updates, 4/2/2011}\\[0.1in]
\noindent {\large\em Niall MacKay}

\noindent Department of Mathematics, University of York, York YO10
5DD, UK \\ email: nm15@york.ac.uk

\vskip 0.2in
\noindent\parbox[t]{5.3in}{\small {ABSTRACT}: We explore the house price distributions for the English cities of London, Manchester, Bristol, Newcastle, Birmingham and Leeds. We find Pareto (power law) behaviour in their upper tails, which is clearly distinct from lognormal and gamma distributions in the cases of London, Manchester and Newcastle. For London, the city with the lowest power, this is a striking match with that found in the wealth distribution of the super-rich. We propose an index of Housing Wealth Inequality based on the Pareto exponent and analogous to the Gini coefficient, and comment on its possible uses.\\[0.1in] {KEY WORDS:} house price distributions, Pareto distribution, housing wealth inequality}

\vspace{0.3in} \noindent{\large\bf  Introduction}

\noindent It is surprising how little attention has been paid until recently to the house price distribution---that is, the frequency distribution of housing by price. However, the topic now seems to be attracting interest. In addition to the studies by McMillen (2007) of Chicago, and by M\"a\"att\"anen and Tervi\"o (2010) of the relationship between US income distribution and house prices, work has also been done on various East Asian locations---Singapore (Han {\em et al.}, 2002), Taiwan (Chou and Li, 2010) and Tokyo (Ohnishi {\em et al.}, 2010). Our intention in this brief note is to use publicly-available data to pique wider interest in this distribution for the English housing market, and above all for that of London.

As Ohnishi {\em et al.\ }note, it is sometimes assumed that the house price distribution is log-normal. That is, if one plots the number of houses $N$ on the market against their price $P$, and in particular if one plots the logarithm of the number, $\log N$, within a price band from $P$ to $(1+\epsilon)P$ (so that such bands are of uniform width on a logarithmic scale), one might expect to see an inverted parabola. Such a distribution is the natural outcome of multiplying random variables, and one might then conclude that high house prices are the product of many random probabilities.

However, it is well known that income and wealth---to which we might expect housing wealth to be closely related---follow a Pareto, power-law distribution, in which the upper tail  (in the log-log plot) is a straight line. Further, the slope of this line is a natural measure of the inequality of the distribution, and tends to be lower for richer groups (Coelho {\em et al.}, 2008). There are various processes which might underlie such a distribution, and a long controversy over the truth they embody---for an introduction see Mitzenmacher (2004).

We shall examine the house price distributions for just over 200,000 properties in six English conurbations: Birmingham (with Wolverhampton), Bristol, Leeds, London, Manchester (with Salford) and Newcastle (with Gateshead and Sunderland). As we shall see, there is a region of power-law behaviour in the upper tail of the distribution of asking house prices in all of these.  In London, Manchester and Newcastle, the cities for which we observed the greatest inequality, lognormal and gamma distributions are rejected for the full upper tail (from the distribution's peak to its upper end). Finally we note the potential utility of extracting a measure of housing wealth inequality from this distribution.

Statistical calculations were performed using $R$.

\vskip 0.2in
\noindent {\bf House prices in English cities}

\noindent We extracted house price distributions for six English conurbations by searching for all properties for sale within a 15 mile radius of a point at their centre. We used the publicly-available data at the property website {\tt home.co.uk}, which combines prospective prices from many different selling agents, to obtain a total of 209,827 prices, divided into 40 logarithmic bins (with $\epsilon=0.16145$). Table 1 gives, for each city, the OSGB grid reference of the centre of the circle, the lower bounding price of each bin, the number in each bin, and the complementary Cumulative Frequency Distribution (CCFD: the cumulative number from the top of the distribution). Figures in Table 1 were gathered between 3rd and 7th December 2010.

At the lower end of the distribution, we observe that properties for sale include plots, leases, garages and so on. For this reason we will not be attempting to impute a curve or process to the lower tail, and we took a base price of \pounds54,950. At the upper end we took as our upper limit the first bin which contained no properties for sale.

\pagebreak
\noindent{\em London}
\begin{figure}[ht!]
\subfigure[natural]{%
\includegraphics[keepaspectratio=true,
width=190pt]{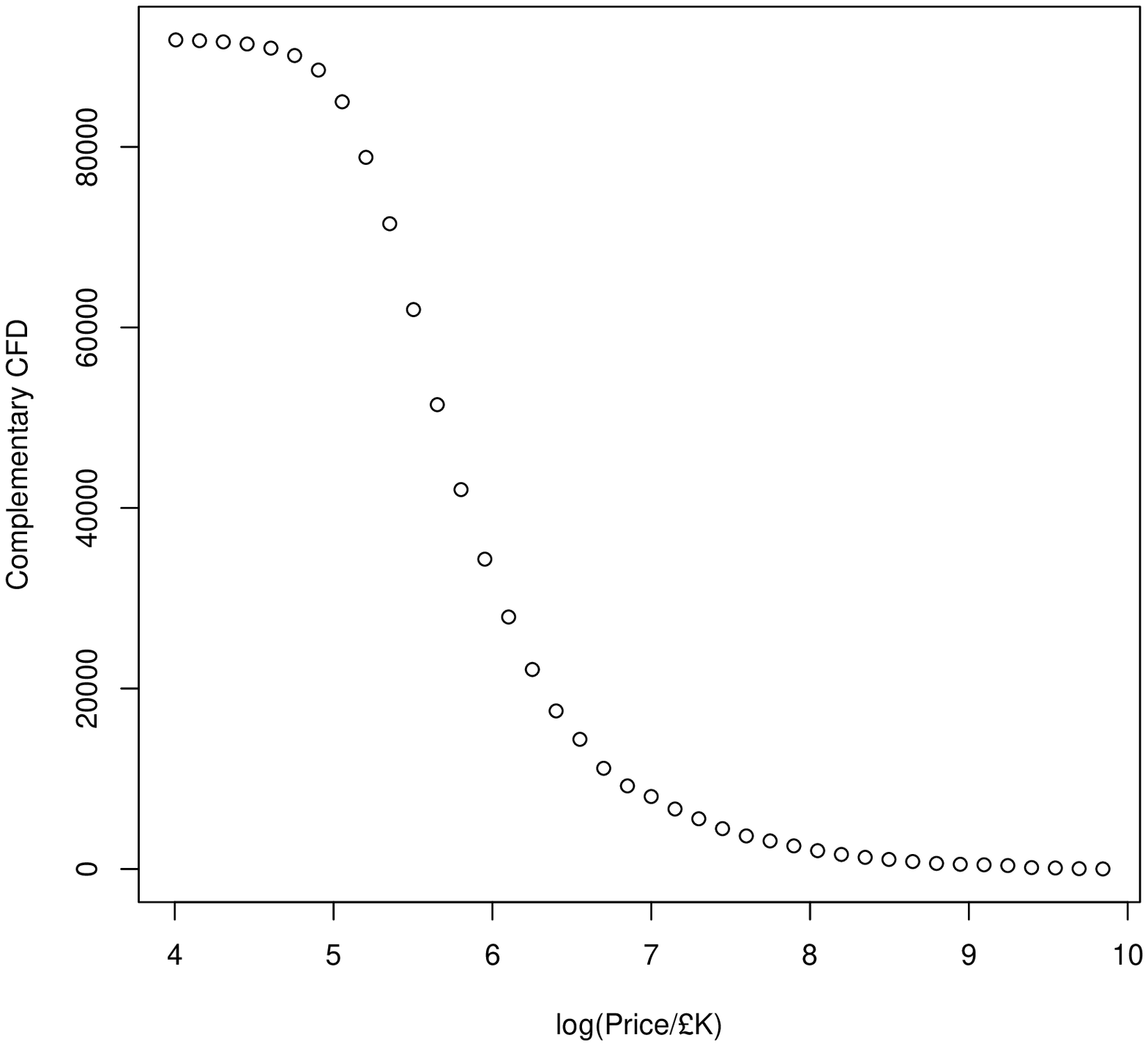}} \hskip 0.6in
\subfigure[logarithm]{%
\includegraphics[keepaspectratio=true,
width=190pt]{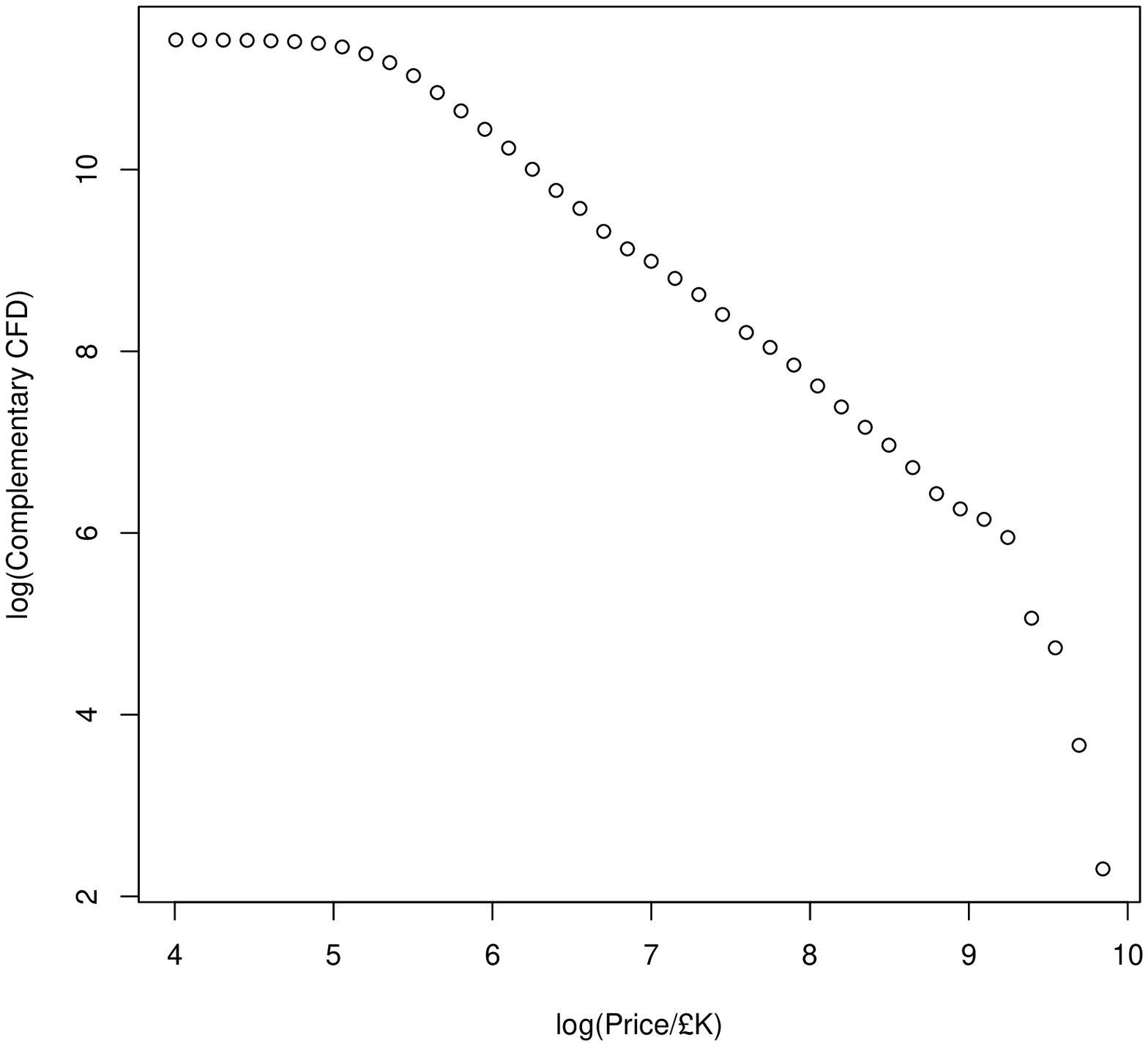}} \caption{Complementary cumulative frequency distribution for London properties}
\end{figure}

\noindent Figure 1 shows the resulting plots for London. All logarithms are natural ({\em i.e.\ }to base $e$). The CCFD is impressively smooth---as they often are---and in its logarithm we observe the striking straight-line behaviour in the tail which is indicative of a power-law distribution.

It is well-known that fitting a power law by linear regression to the logarithm of the FD (the Frequency Distribution:  minus the gradient of the CCFD) is an unsafe procedure, mostly owing to uneven variation in the tail (Goldstein {\em et al.}, 2004), but it will nevertheless be worth examining it carefully, and we plot it, and its logarithm, in Figure 2.

\begin{figure}[ht!]
\subfigure[natural]{%
\includegraphics[keepaspectratio=true,
width=190pt]{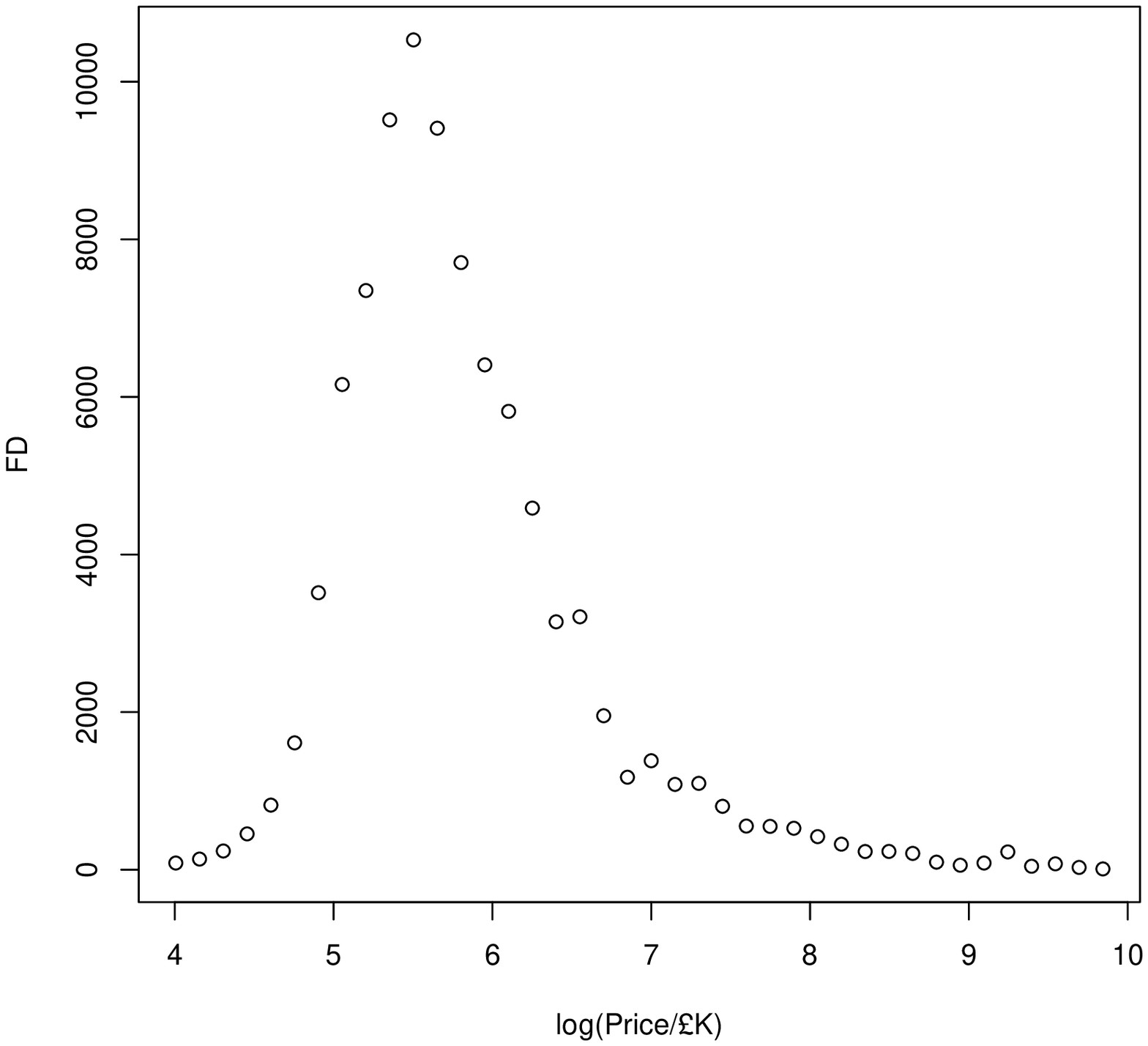}} \hskip 0.6in
\subfigure[logarithm]{%
\includegraphics[keepaspectratio=true,
width=190pt]{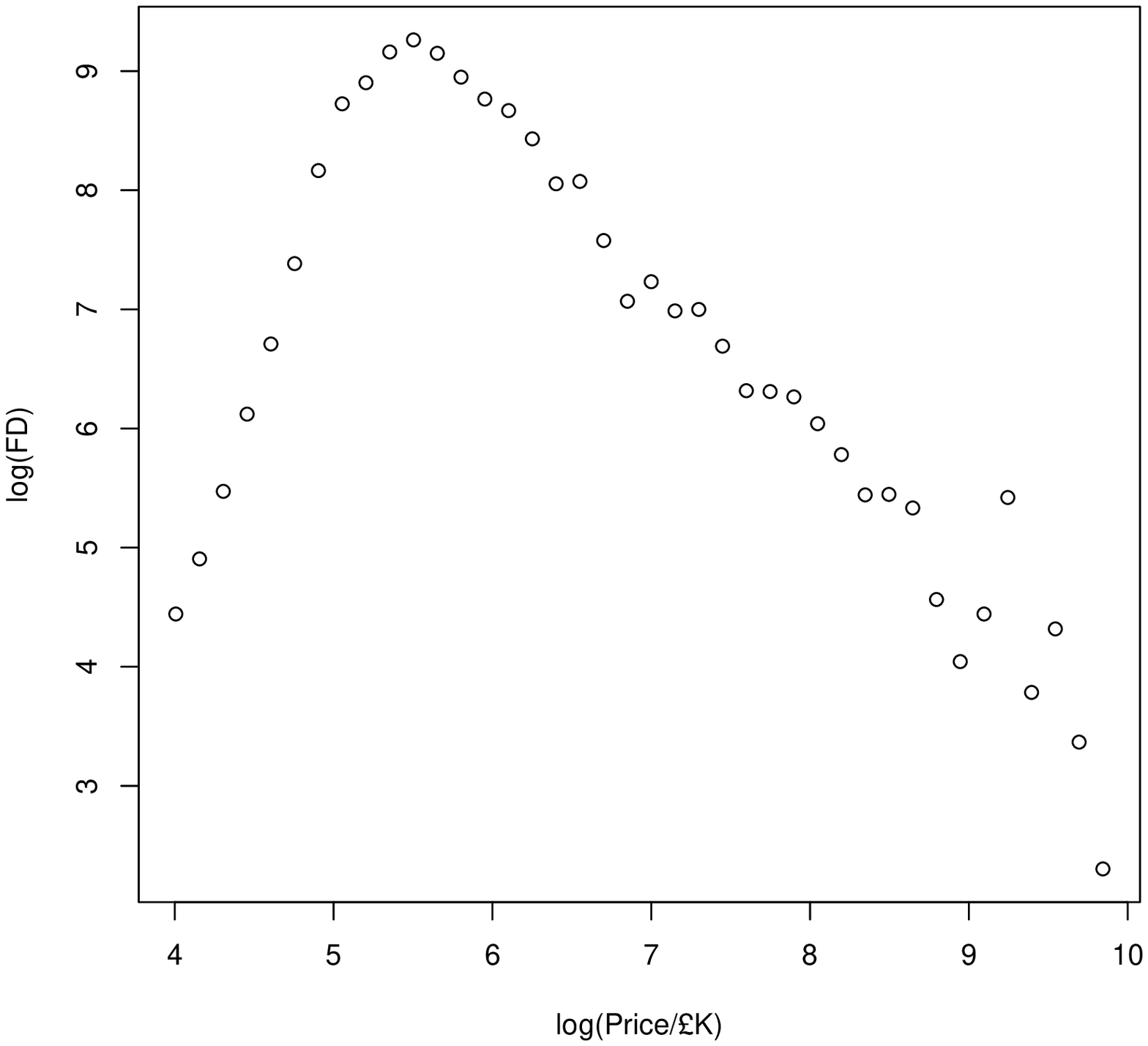}} \caption{Frequency distribution for London properties}
\end{figure}

A Pareto or power-law distribution with exponent $\alpha$ has CCFD
\be\label{CCFD}
F(P) = f_{\rm M}\left({P\over P_{\rm M}}\right)^{-\alpha}\,,\ee
where $P_{\rm M}$ is the maximum price, occurring with frequency $f_{\rm M}$. The FD is therefore
\be\label{FD} f(P)=-F'(P)= \alpha f_{\rm M} P^{-\alpha-1} P_{\rm M}^\alpha.\ee
Thus with natural frequency (additive) bins, the FD has an exponent one lower than the CCFD. It is worth noting, however, that with our  logarithmic (multiplicative) bins $d(\log P)=dP/P$ one observes frequencies
\be\label{dFD} f(P) \, dP = \alpha f_{\rm M} \left({P\over P_{\rm M}}\right)^{-\alpha} \,d(\log P)\ee
and thus the same exponent in FD and CCFD.
Typically $\alpha>1$, and indeed the mean (respectively the $n$th moment) diverges as $P\rightarrow\infty$ if $\alpha\leq1$ (respectively $\alpha\leq n$). The FD and CCFD also diverge as $P\rightarrow 0$, so one usually imposes a minimum-price cut-off $P_{\rm m}$, so that $P_{\rm m}\leq P \leq P_{\rm M}$.

It is clear that there is some randomness at the very top of the distribution, where the frequencies are small and the bins large, and we might expect psychological price barriers to be especially important. Rather than use maximum likelihood estimators, our strategy was to use least squares regression on the CCFD from its peak up to $P_{\rm M}$. We then analyze the fit for outliers, successively removing data points from the top until none has a Cook's distance $D>1$. (The Cook's distance $D$ of a data point measures the extent to which it has skewed the outcome; with this strategy all our data, for all cities, then had Cook's distances $D<0.7$.)

As one might guess merely from observation of the CCFD plot (Figure 1), for London this led us to remove the top four points, leaving points 11-36, corresponding to a price range from just under \pounds250K to just over \pounds10M. These gave $\alpha=-1.37\pm0.01$, with a multiple-$R^2>0.999$. (Notice that in the FD plot, Figure 2, there are four more outlying points---if we had removed these too we would have found $\alpha=1.35\pm0.01$.)

It is perhaps also worth noting that UK property purchase tax (`Stamp Duty') thresholds are at \pounds250K (bin 11) and \pounds500K (bin 15). Slightly to our surprise (since these taxes rise at each threshold to a higher proportion of the {\em full} purchase price), there was no obvious significant deviation at these points. Two other points are outliers on visual inspection, bin 18 (high) and bin 20 (low), but these did not significantly affect the fitted $\alpha$.

One worthwhile check of robustness was to shift the centre of the search (we tried TQ304816 and TQ277808), but this had negligible effect. Further, the same search two months earlier (7th October 2010) yielded a near-identical distribution and $\alpha=1.37\pm0.01$.

\vskip 0.1in
\noindent{\em Other cities}

\noindent We begin with the plots of log(CCFD), Figure 3, and log(FD), Figure 4, for the five other cities.

\begin{figure}[ht!]
\subfigure[Manchester]{%
\includegraphics[keepaspectratio=true,
width=190pt]{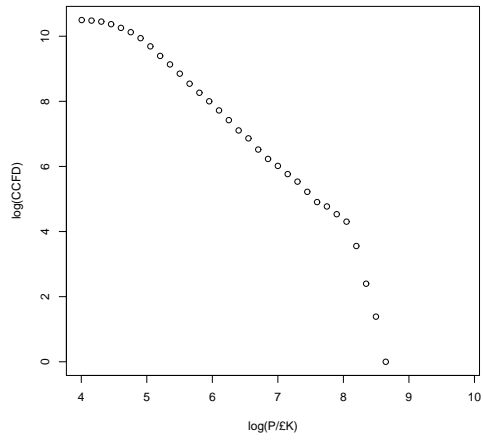}} \hskip 0.6in
\subfigure[Bristol]{%
\includegraphics[keepaspectratio=true,
width=190pt]{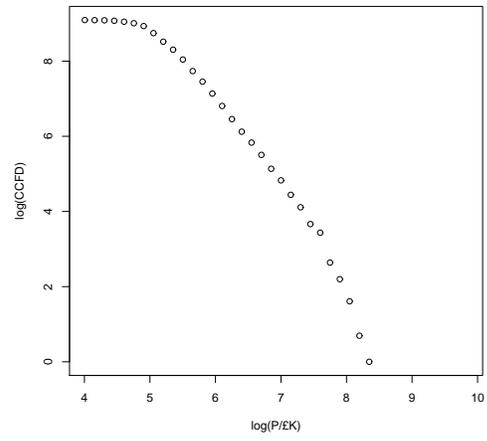}} \\
\centerline{\subfigure[Newcastle]{%
\includegraphics[keepaspectratio=true,
width=190pt]{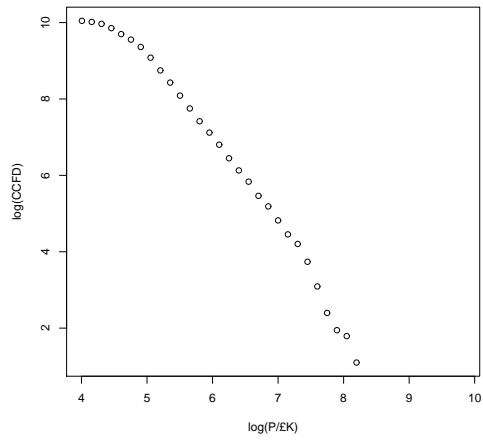}} }\\
\subfigure[Birmingham]{%
\includegraphics[keepaspectratio=true,
width=190pt]{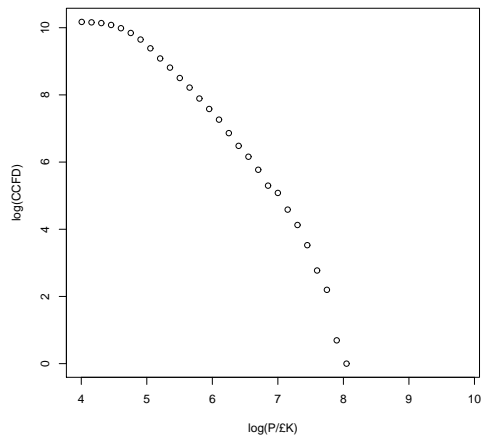}} \hskip 0.6in
\subfigure[Leeds]{%
\includegraphics[keepaspectratio=true,
width=190pt]{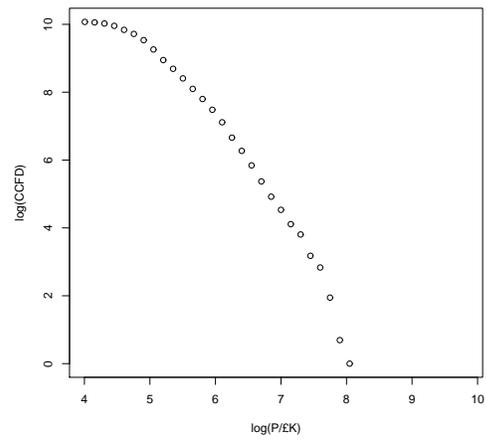}}
\caption{Cumulative frequency distributions for property prices in English cities}
\end{figure}

\begin{figure}[ht!]
\subfigure[Manchester]{%
\includegraphics[keepaspectratio=true,
width=190pt]{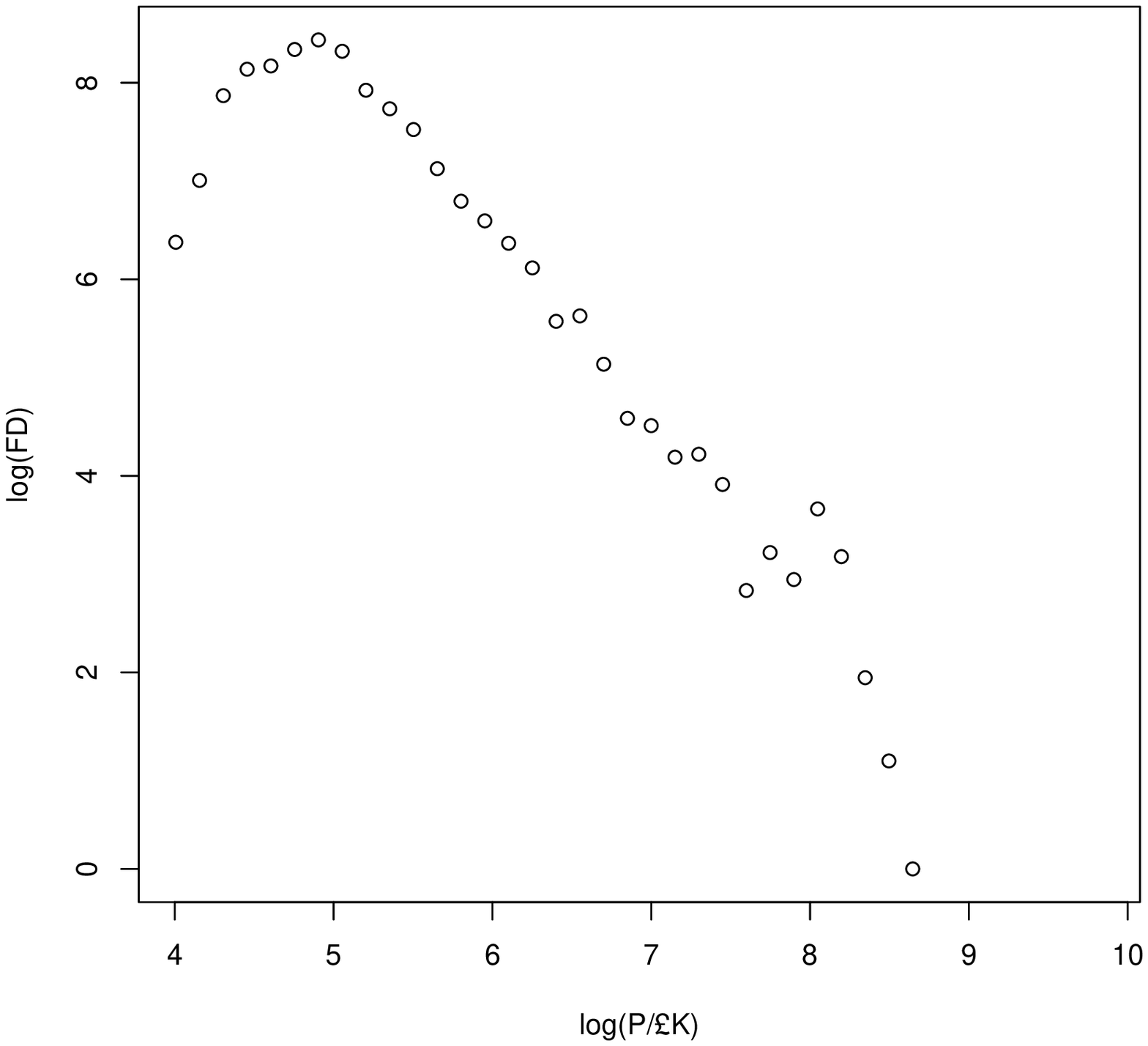}} \hskip 0.6in
\subfigure[Bristol]{%
\includegraphics[keepaspectratio=true,
width=190pt]{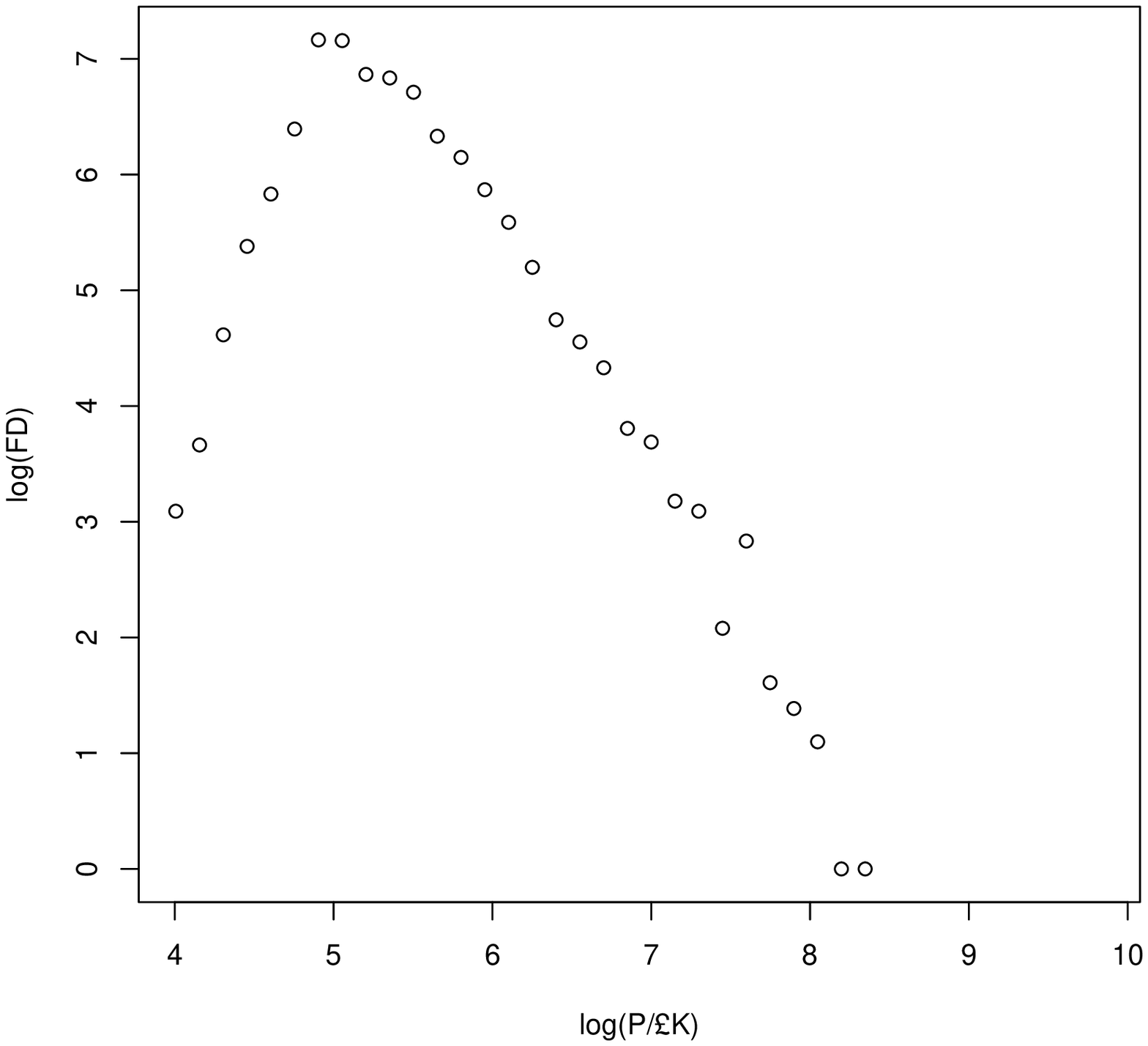}} \\
\centerline{\subfigure[Newcastle]{%
\includegraphics[keepaspectratio=true,
width=190pt]{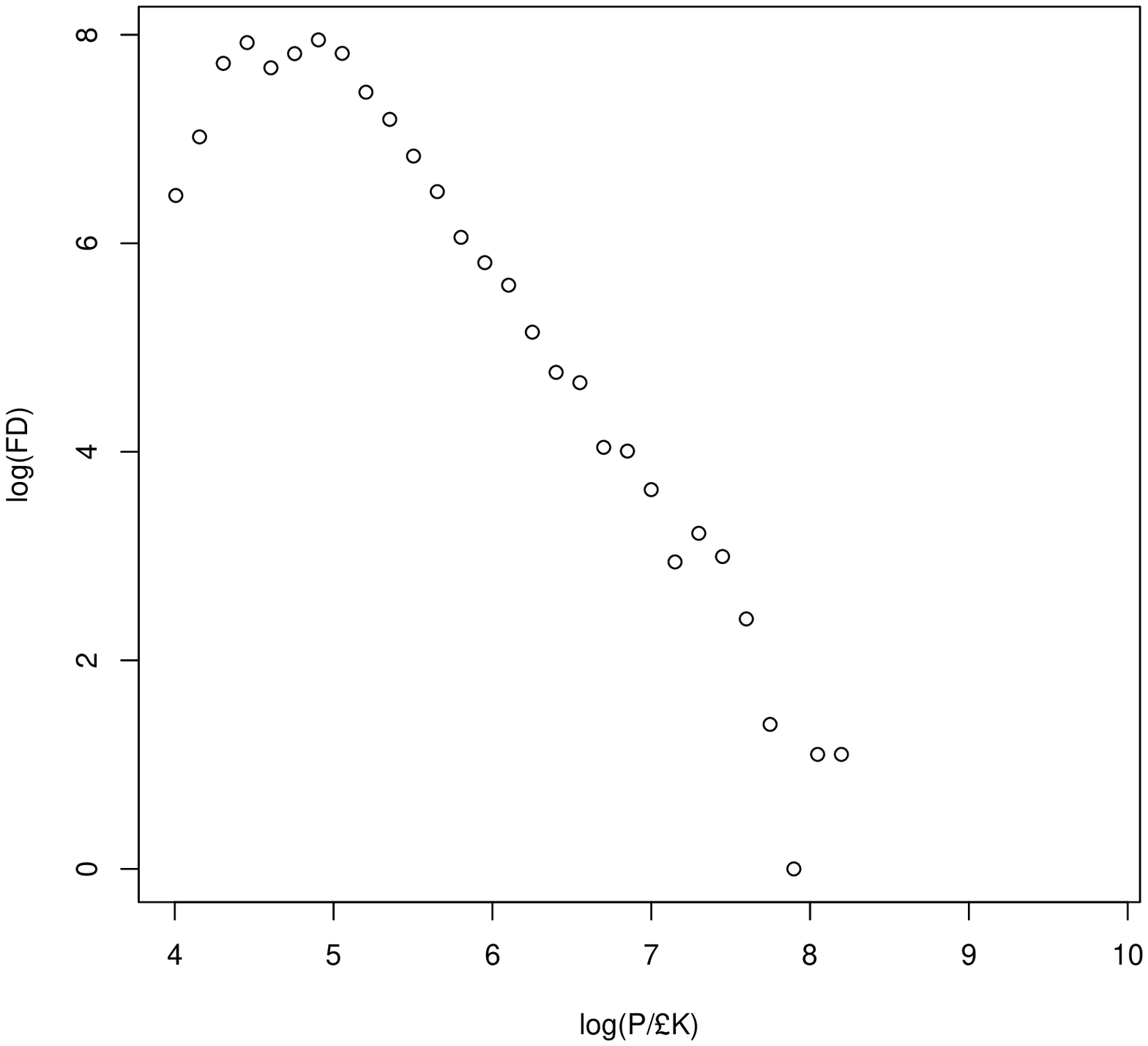}} }\\
\subfigure[Birmingham]{%
\includegraphics[keepaspectratio=true,
width=190pt]{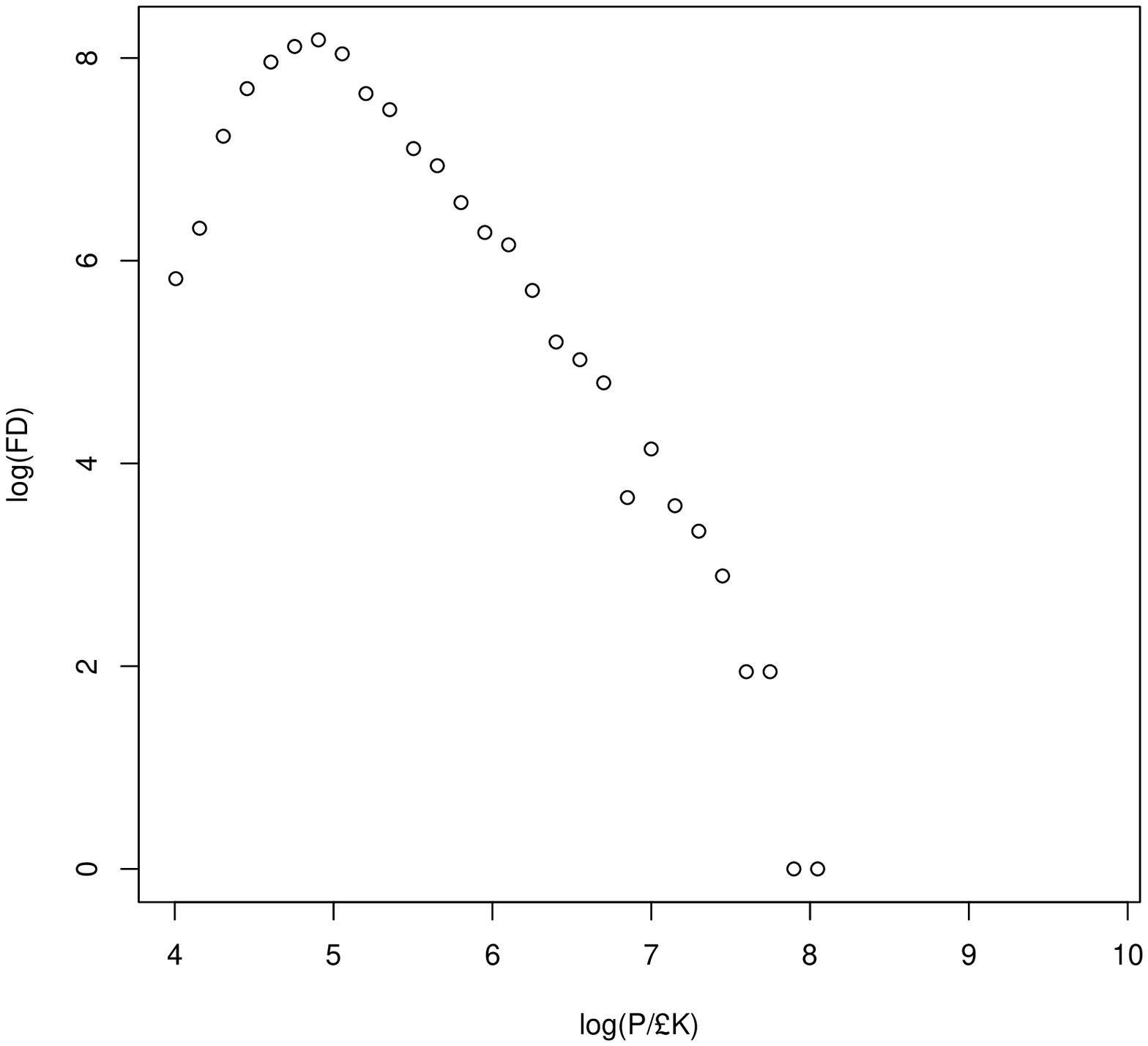}} \hskip 0.6in
\subfigure[Leeds]{%
\includegraphics[keepaspectratio=true,
width=190pt]{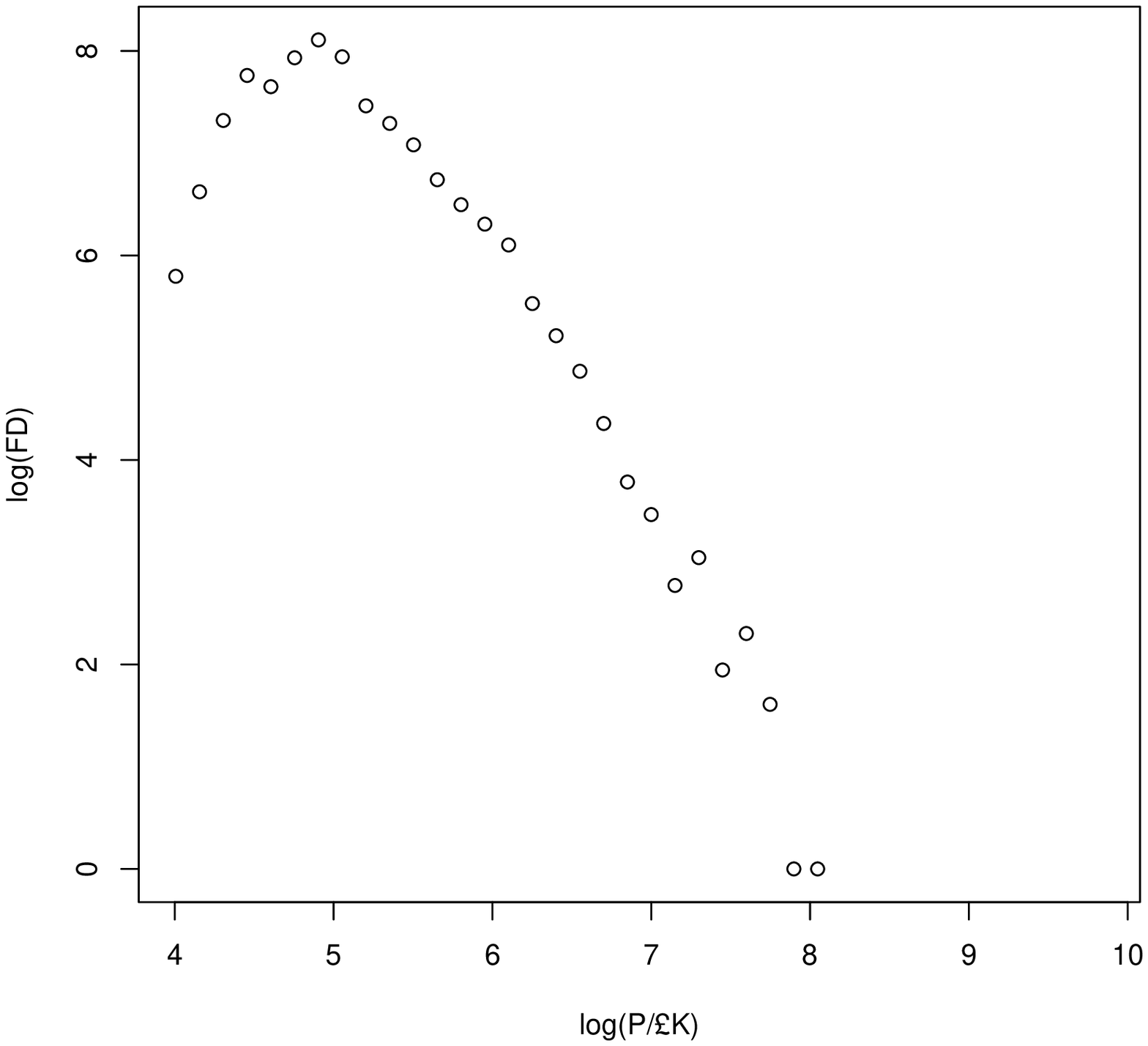}}
\caption{Frequency distributions for property prices in English cities}
\end{figure}

It is natural to suggest psychological explanations for some of the minor anomalies. For example, in Newcastle, Leeds and Manchester we note above-trend numbers in bin 4 (just under \pounds100K) and below in bin 5 (just over \pounds100K). Effects at the tax thresholds do not appear to be strong. As with London we note that bin 20 appears anomalously low in Birmingham, and slightly so in Manchester and Bristol---this bin is around the \pounds1M threshold.

We followed the same algorithm as for London to fit a power-law to the tail. Table 2 gives, for each city, the peak and highest non-empty bins, $\alpha$ and its standard error, and the excluded outliers. (We also give $p_1$, $p_2$ and HWI, all to be defined below.)

We chose regression on the log(CCFD) rather than using maximum likelihood estimators because we wanted a transparent, algorithmic means of treating the outliers at the top of the distribution. However, an upshot is that the standard errors are probably underestimates by a factor of two or so, and that $\alpha$ should be considered accurate only in its first decimal place. (For example, we noted earlier that, had we removed 8 rather than 4 outliers from the London data, the value of $\alpha$ would have shifted by $0.02$.)

We should ask whether these distributions are truly power laws, or whether the drop-off evident at the top of the distribution vitiates these. We did so by testing two alternative hypotheses: first, that the distribution might be log-normal (and thus an inverted parabola in the log(FD) plots); and, second, that it might approximate a gamma distribution, in which (\ref{FD},\ref{dFD}) include an extra factor $\beta^P$ for some $\beta\simeq 1$. One should note that although in both cases scale-invariance is lost, thereby introducing a price-scale, the significance of the departure from linearity (respectively quadratic and exponential) is invariant under changes of this scale.

As we noted earlier, we do not believe that the data at the bottom of the distribution are susceptible to a single, unified explanation with those at the middle and top. Thus it would be inappropriate to seek, for example, an exotic distribution such as the double Pareto-lognormal (Reed and Jorgensen, 2004) to explain all data. This would appear in the log-log FD plot essentially as two straight-line sections with an interpolating curve. Similarly one should not take the lower tail of the distribution as evidence against a power-law in the upper.

Our approach for testing log-normal (we quote $t$-test $p_1$) and gamma ($p_2$) distributions, therefore, was to fit linear models  respectively quadratic and exponential in $\log P$ to the same points to which we previously fitted a straight line, from the peak to the top. For London, Manchester and Newcastle, by this means we reject (both $p_1>0.05$ and $p_2>0.05$) log-normal and gamma behaviour in prices from \pounds1M upwards for London, and \pounds330K upwards for Manchester and Newcastle. For the other cities no such rejection was warranted. For example, one might consider that for Leeds and Birmingham there is some curvature in the upper tail beyond the mere sparseness of properties at very high prices. This curvature is of the correct sign and is significant at $p<0.05$, although it cannot be estimated accurately enough to imply a mean.

\vskip 0.2in\noindent{\bf An index of housing wealth inequality?}

\noindent Pareto exponents tend to be lower among richer societies (Coelho, 2008), and our lowest value, for London, of $\alpha=1.37$ is a striking match with the Forbes magazine data for the world's richest individuals (as reported, for example, in Richmond {\em et al.}, 2006), which had $\alpha=1.37$ (in 2006) and $\alpha=1.38$ (2003).

It therefore seems to us that it would be interesting to use properties of the house price distribution as proxies for similar properties of the general wealth distribution. As a measure of inequality one might use the Gini coefficient $G$, one minus the ratio of the sum of individuals' ranks (from least-wealthy upwards) multiplied by their wealth as a proportion of its maximum possible value (so that $0\leq G \leq 1$). This utilizes the whole distribution, of course. Because of what we believe to be the paucity of meaning at the top and bottom of our distributions, we would propose instead using a function of $\alpha$. For example, one might propose simply its inverse, $I:=1/\alpha$.

However, if one computes $G$ for a pure Pareto disstribution, one obtains $G={1\over 2\alpha-1}$ (a straightforward calculation, or see Dorfman, 1979). Thus we propose rather {\em defining} our index of Housing Wealth Inequality to be HWI$:={1\over 2\alpha-1}$ for the $\alpha$ computed earlier---and in fact this differs little from the $I=1/\alpha$ suggested above, with HWI$(I)$ obeying HWI$(0)=0$, HWI$(1)=1$ and max$|$HWI$-I|\simeq 0.17$.

Such an index is likely to be more robust than, for example, the ratio of standard deviation to mean, used in Van Nieuwerburgh and Weill (2006), which potentially suffers from the problem mentioned earlier: that in a pure Pareto distribution with $\alpha<2$  the variance (and, for $\alpha<1$, also the mean) is divergent. This measure is likely to be skewed by the top of the upper tail.

The HWI has various possible uses, and certain advantages. Clearly it combines, via the devotion to property ownership of English society, aspects of the distributions of both income and wealth. If used alongside a measure of average wealth, perhaps the median house price, it might assist in disentangling average wealth from wealth inequality. Above all, it is easily tied to a particular locale: one can measure, quickly and easily, such properties for any town, city or larger region. For our six cities, HWI was highest for London, followed by Manchester, the two cities for which the Pareto distribution was most clear. Among the other four the HWI is less clearly resolved, although it is clearly greater for Newcastle than for Leeds and Bristol.

Again we defer the calculation of time-series data for HWI, and a search for its correlates, to future work, for which we would need to purchase historical data from the UK Land Registry. However, we note that, because of its ease of computation for a geographical area, it might be particularly useful in investigating one of the more interesting (and controversial) social theses of recent years: that various social and public-health outcomes are more positive for societies and communities which have lower indices of inequality, independent of their overall wealth (Wilkinson and Pickett, 2009).

\vskip 0.1in
\noindent{\bf Acknowledgments}\\
I should like to thank Kate Pickett and Sonia Mazzi for discussions.

\parindent 0pt
\parskip 6pt
\baselineskip 16pt

\pagebreak
{\large\bf References}

Coelho, R.,  Richmond, P., Barry, J. \& Hutzler, S. (2008) Double power laws in income and wealth distributions. {\em Physica A} 387, pp.\ 3847-3851

Chou, C.-I. \& Li, S. P. (2010)
House price distributions of Taiwan: a preliminary study.
arXiv:1008.1376v1 [physics.soc-ph]

Dorfman, R. A formula for the Gini coefficient. (1979) {\em The Review of Economics and Statistics} 61(1), pp.\ 146-149

Goldstein, M. L., Morris, S. A. \& Yen G. G. (2004) Problems with fittting to the lower-law distribution.
{\em European Physical Journal} 41(2), pp.\ 255-258

Han, S. S., Yu, S. M., Malone-Lee, L. C. \& Basuki, A. (2002)
Dynamics of property value distribution in an Asian metropolis: the case of landed housing in Singapore, 1991-2000. {\em Journal of Property Investment and Finance} 20, pp.\ 254-276

M\"a\"att\"anen, N. \&  Tervi\"o, M. (2010) Income distribution and housing prices: an assignment model approach.
CEPR Discussion Papers no.\ 7945

McMillen, D. P. (2007) Changes in the distribution of house prices over time: Structural characteristics, neighborhood, or coefficients? {\em Journal of Urban Economics} 64(3), pp.\ 573-589

Mitzenmacher, M. (2004) A brief history of generative models for power law and lognormal distributions. {\em Internet Mathematics} 1(2), pp.\ 226-251

Ohnishi, T., Mizuno, T., Shimizu, C. and Watanabe, T. (2010)
On the evolution of the house price distribution. {\em Understanding Inflation Dynamics of the Japanese Economy}, working paper series no.\ 56.

Reed, W. J. \& Jorgensen M. (2004) The double Pareto-lognormal distribution---a new parametric model for size distributions. {\em Communications in Statistics} 33(8), pp.\ 1733-1753

Richmond, P., Hutzler, S., Coelho, R. \& Repetowicz, P. (2006) A review of empirical studies and models of income distributions in society. {\em Econophysics and Sociophysics: trends and perspectives}, eds Chakrabarti, B. K., Chakraborti, A. \& Chatterjee A. (Weinheim: Wiley)

Van Nieuwerburgh, S. \& Weill, P.-O. (2006) Why has house price dispersion gone up?
NBVER working paper 12538.

Wilkinson, R. \& Pickett, K. (2009) {\em The Spirit Level: Why More Equal Societies Almost Always Do Better} (London: Allen Lane)

\pagebreak
\vspace*{-0.6in}
\centerline{Table 1: Distributions of house prices in six English cities}

\rotatebox{90}{ {\small
\begin{tabular}{|c|c|cc|cc|cc|cc|cc|cc|}
\\[-0.6in]\hline
N &	P	&	London	& &	Manchester &&	Bristol	&& Newcastle && Birmingham &&	Leeds & \\
(Centre:) && TQ 292 772 && SJ 832 987 && ST 700 721 && NZ 257 640 && SP 069 869 && SE 297 338 & \\
1 &	\pounds54.95K &		85 &	91858	& 588&	36183&	22&	8893&	638&	23073&	338&	26167&	 329&	 23653 \\
2&	63.83	&	135&	91773&	1103&	35595&	39&	8871&	1119&	22435&	556&	25829&	752&	23324\\
3&	74.13	&	238&	91639&	2614&	34492	& 101	& 8832	& 2266	& 21316	& 1378	& 25273	& 1511 &	 22572\\
4&	86.10	&	455	&91400&	3421&	31878&	217	&8731&	2764&	19050	&2204&	23895&	2346&	21061\\
5&	100.00	&	820	&90945	&3538&	28457&	341	&8514&	2171&	16286&	2865&	21691	&2102&	18715\\
6&	116.14	&	1610&	90125&	4182&	24919&	598	&8173&	2487&	14115&	3340	&18826&	2789	&16613\\
7&	134.90	&	3515&	88515&	4610&	20737&	1292	&7575	&2837&	11628	&3564	&15486&	3322	& 13824\\
8&	156.68	&	6158&	85000&	4106&	16127&	1284&	6283&	2493&	8791&	3106&	11922&	2816&	 10502\\
9&	181.97	&	7350&	78842&	2762&	12021&	959	&4999	&1718	&6298	&2099	&8816	&1743&	7686\\
10&	211.35	&	9516&	71492&	2287&	9259&	930&	4040&	1324&	4580&	1791&	6717&	1468&	5943\\
11&	245.47	&	10531&	61976&	1851&	6972&	822&	3110&	931&	3256&	1220&	4926&	1190&	4475\\
12&	285.10	&	9410	&51445&	1243&	5121&	562&	2288&	662&	2325&	1030&	3706&	846&	3285\\
13&	331.13	&	7705&	42035	&893&	3878&	468&	1726&	427&	1663&	716&	2676&	663&	2439\\
14&	384.59	&	6407&	34330&	731	&2985&	354	&1258&	335&	1236&	533&	1960&	548&	1776\\
15&	446.68	&	5817&	27923&	582	&2254&	267&	904&	270&	901&	472&	1427&	447&	1228\\
16&	518.80	&	4589&	22106&	453	&1672&	181	&637	&172	&631	&301	&955	&252	&781\\
17&	602.56	&	3146&	17517&	263&	1219&	115	&456	&117	&459	&181	&654	&184	&529\\
18&	699.84	&	3210&	14371&	278	&956&	95&	341&	106&	342&	152&	473&	130&	345\\
19&	812.83	&	1954&	11161&	170&	678&	76&	246&	57&	236&	121&	321&	78&	215\\
20&	944.06	&	1174&	9207	&98	&508&	45	&170&	55&	179&	39	&200&	44&	137\\
21&	1096.48	&	1383&	8033&	91	&410&	40	&125	&38&	124&	63&	161&	32	&93\\
22&	1273.50	&	1083&	6650&	66	&319&	24&	85&	19&	86&	36&	98&	16&	61\\
23&	1479.11	&	1096&	5567&	68	&253&	22&	61&	25&	67&	28&	62&	21&	45\\
24&	1717.91	&	805&	4471&	50	&185&	8&	39&	20&	42&	18&	34&	7	&24\\
25&	1995.26	&	554&	3666&	17	&135&	17&	31&	11&	22&	7&	16&	10&	17\\
26&	2317.39	&	550	&3112&	25	&118	&5	&14	&4	&11&	7&	9&	5&	7\\
27&	2691.53	&	526& 	2562&	19&	93&	4&	9&	1&	7&	1&	2&	1&	2\\
28&	3126.08	&	420	&  2036&	39&	74&	3&	5&	3&	6&	1&	1&	1&	1\\
29&	3630.78	&	324&	1616&	24&	35&	1&	2&	3&	3	&&&&\\
30&	4216.97	&	231&	1292&	7&	11&	1&	1	&&&&&&\\
31&	4897.79	&	232&	1061&	3&	4	&&&&&&&&\\
32&	5688.53	&	207&	829&	1&	1	&&&&&&&&\\
33&	6606.93	&	96&	622	&&&&&&&&&&\\
34&	7673.61	&	57&	526	&&&&&&&&&&\\
35&	8912.51	&	85&	469	&&&&&&&&&&\\
36&	10351.42&	226&	384	&&&&&&&&&&\\
37&	12022.64&	44&	158	&&&&&&&&&&\\
38&	13963.68&	75&	114	&&&&&&&&&&\\
39&	16218.10&	29&	39	&&&&&&&&&&\\
40&	18836.49&	10&	10 &&&&&&&&&&\\
\hline
\end{tabular}
}}

\vskip 0.15in
\centerline{Table 2: Fits for English cities}

\hspace*{-0.2in}
\begin{tabular}{|c|ccccccc|}
\hline
City & Peak & Last non-zero & $\alpha$ & Removals & $p_1$ & $p_2$ &HWI \\ \hline
London & 11 & 40 & $1.37\pm0.01$ & 37-40 &0.397 & 0.136 & $0.57\pm0.001$ \\
Manchester & 7 & 32 & $1.83\pm0.01$ & 29-32 & 0.214 & 0.0528 & $0.38\pm0.001$ \\
Bristol & 8 & 30 & $2.50\pm0.09$ & none & $10^{-4}$ & $10^{-4}$ & $0.25\pm0.01$\\
Newcastle & 7 & 29 & $2.17\pm0.05$ & 27 & 0.274 & 0.300 & $0.30\pm0.01$\\
Birmingham &7 & 28 & $2.29\pm0.04$ & 24-28 & $10^{-4}$ & $10^{-5}$ & $0.28\pm 0.01$\\
Leeds & 7 & 28 & $2.53\pm0.05$ & 26-28 & $10^{-4}$ & $10^{-4}$ & $0.25\pm0.01$\\
\hline
\end{tabular}

\end{document}